\documentclass[reprint,aps,twocolumn,showpacs,superscriptaddress,groupedaddress,nofootinbib]{revtex4-2}
\usepackage{graphicx} 
\usepackage{tabularx}
\usepackage{booktabs}
\usepackage{dcolumn}   
\usepackage{bm}        
\usepackage{amssymb}   
\usepackage{amsmath}   
\usepackage{multirow}
\usepackage{hyperref}
\usepackage{braket}
\usepackage{subfigure}
\usepackage{array}
\usepackage[math]{cellspace}
\usepackage[utf8]{inputenc}
\usepackage{cleveref}
\usepackage{algorithm}
\usepackage{algpseudocode}
\usepackage{physics}
\usepackage{url}
\hypersetup{
bookmarksnumbered,
colorlinks=true,
linkcolor=magenta,
urlcolor=magenta,
citecolor=blue
}
\begin{document}

 \title{Chiral and isospin breaking in the two-flavor Schwinger model}

\date{\today}

\author{D. Albandea}
\email{david.albandea@uv.es}
\affiliation{Instituto de Física Corpuscular (CSIC-UVEG), Parc Cient\'{\i}fic de la Universitat de Val\`encia ,
C/ Catedr\'atico Jos\'e Beltr\'an 2, E-46980 Paterna, Spain}
\author{P. Hern\'andez}
\email{m.pilar.hernandez@uv.es}
\affiliation{Instituto de Física Corpuscular (CSIC-UVEG), Parc Cient\'{\i}fic de la Universitat de Val\`encia ,
C/ Catedr\'atico Jos\'e Beltr\'an 2, E-46980 Paterna, Spain}

\begin{abstract}
The Schwinger model with two massive fermions is a nontrivial theory for which
no analytical solution is known. The strong coupling limit of the theory allows for different semiclassical approximations to extract
properties of its low-lying spectrum. In particular, analytical
results exist for the fermion condensate, the fermion mass dependence of the pseudoscalar meson mass or its decay constant. These approximations, nonetheless, are not able to quantitatively
predict isospin breaking effects in the light spectrum, for example. In this paper we use lattice simulations to test various analytical
predictions, and study isospin
breaking effects from nondegenerate quark masses.
We also introduce a low-energy effective field theory based on a
nonlinear $\sigma$ model with a dilaton field, which leads to the correct fermion mass dependence of the pion mass, the correct $\sigma$-to-$\pi$ mass ratio and a prediction of the isospin breaking effects, which we test numerically.
\end{abstract}
\maketitle

\section{Introduction}

Two-dimensional quantum field theories have historically provided valuable
insights into nonperturbative phenomena in more complex systems. One such theory
is the Schwinger model and its variants.
The Schwinger model is a theory of a massless fermion coupled to a $\mathrm{U}(1)$ gauge field
\cite{Schwinger:1962tp}, which can be nontrivially extended with $N_{f}$ fermion
flavors, as well as with fermion masses. A plethora of nonperturbative
phenomena analogous to those expected in non-Abelian gauge theories in four
dimensions (4D) are
present in these simpler theories~\cite{Coleman:1975pw}, including confinement,
fermion condensation and chiral symmetry breaking from anomalies.

The Schwinger model with $N_f$  massless fermions is exactly solvable and trivial. In the
case of $N_f=1$, the theory reduces to that of a free massive scalar
singlet~\cite{Schwinger:1962tp}. Interestingly, the Witten-Veneziano relation~\cite{Witten1979,Veneziano1979}
between the mass of this heavy boson (analogous to the $\eta'$ in QCD)  and the
topological susceptibility in the quenched theory is exact in the
Schwinger model~\cite{Seiler:1987ig,Giusti:2001xh}, as it does not rely on any
large $N_c$ limit---as happens in QCD.

For $N_f>1$, the theory possesses
a $\mathrm{U}(N_{f})_{L} \times \mathrm{U}(N_{f})_{R}$ flavor symmetry at
the classical level and shows critical behavior. The standard study of the theory
through
bosonization~\cite{Coleman:1976uz} reveals the presence of a massive scalar
sector and a massless, conformal one. This is surprising, since the
Mermin-Wagner theorem forbids spontaneous symmetry breaking in
2D~\cite{Coleman:1973ci}: the full flavor group $\mathrm{SU}(N_f)_L \times \mathrm{SU}(N_f)_R$
remains unbroken and therefore no Goldstone bosons are expected. This model has
been recently studied as an example of ``unparticle'' physics~\cite{Georgi:2020jik,Georgi:2007ek}.

In the presence of fermion masses, no exact solution is known for any $N_f$. The
bosonized model has been studied by semiclassical
methods~\cite{Dashen:1974ci,Coleman:1976uz,Belvedere:1978fj, Hetrick:1995wq}.
For $N_f=2$, and in the strong coupling limit, the theory
reduces to a sine-Gordon model, whose full scattering matrix is known
 \cite{Zamolodchikov:1995xk}. From this exact solution, predictions
such as the fermion mass dependence of the spectrum, the fermion condensate or
the axial current matrix element can be derived~\cite{Smilga:1996pi}. An
interesting observation is that, since the Witten-Veneziano relation is exact in
this model and the topological susceptibility for $\mathrm{U}(1)$ in 2D is known
analytically~\cite{Seiler:1987ig,Gattringer:1993ec,Giusti:2001xh}, a prediction
for the matrix element of the singlet axial current (analogous to $F_{\eta'}$ in
QCD) follows, but, as we will see, it differs from the exact prediction of the
nonsinglet axial current matrix element (analogous to $F_\pi$ in QCD).

In the case of nondegenerate fermion masses, the bosonized theory shows that
there is no isospin breaking in the strong coupling limit~\cite{Coleman:1976uz}.
This fact has been recently revisited in Ref.~\cite{Georgi:2020jik} and
explained in terms of the phenomenon of conformal coalescence in unparticle
physics~\cite{Georgi:2007ek}. The concept of ``automatic
fine-tuning'' is introduced: isospin symmetry breaking at the Lagrangian level leads to
effective isospin symmetry in the low-energy spectrum up to {\it exponentially
suppressed} corrections. No analytical prediction exists for isospin breaking
corrections, since they vanish in the strong coupling limit.

A few numerical studies of the lattice discretized $N_f=2$ model can be found in
the literature
\cite{Dilger:1995pi,Gattringer:1999gt,Gutsfeld:1999pu,Castellanos:2022vic,Funcke:2023lli,Itou:2023img,Castellanos:2023rdw,Chreim:2023tvs,Chreim:2023vmf},
however no conclusive comparison with the exact predictions of the strong
coupling limit versus the semiclassical ones has been obtained. In fact, recent
works have reported deviations from the exact predictions
\cite{Castellanos:2023rdw}. The first study of isospin breaking corrections was
done recently~\cite{Chreim:2023vmf, Chreim:2023tvs} with inconclusive results.

In this work we present the result of a new numerical study\footnote{The code used
    for the simulations and analysis can be found at
    \url{https://github.com/dalbandea/LFTU1.jl}~\cite{https://doi.org/10.5281/zenodo.15051400} and \url{https://github.com/dalbandea/LFTsAnaTools.jl}
.} of the
lattice $N_f=2$ Schwinger model with Wilson fermions. In the degenerate case, we
study the pseudoscalar meson masses, the axial current and pseudoscalar
density matrix elements as a function of the fermion masses, and compare them
with the semiclassical and exact predictions in the strong coupling limit. We
also study isospin breaking corrections in the spectrum in the presence of
nondegenerate fermion masses. Furthermore, we introduce a low-energy effective
theory based on a nonlinear $\sigma$ model with a dilaton field that correctly
reproduces the exact results and gives a parameter-free prediction for the
isospin breaking corrections in the pseudoscalar meson spectrum. We compare this
prediction to our numerical results and discuss the relation of our findings
with the concept of automatic fine-tuning.

The paper is organized as follows. In Secs.~\ref{sec:SM} and \ref{sec:WI} we
review the known analytical predictions for the two-flavor Schwinger model and
the chiral Ward identities in the absence of spontaneous chiral symmetry
breaking, respectively. In Sec.~\ref{sec:EFT} we introduce a low-energy
effective theory, based on a nonlinear $\sigma$ model with a dilaton field, that
describes the dynamics of the lightest degrees of freedom: the scalar singlet
and the triplet of pseudoscalar mesons. We also add the pseudoscalar singlet
through the $\mathrm{U}(1)_A$ anomaly and derive an exact prediction of the isospin
breaking corrections in the pseudoscalar meson spectrum. In
Sec.~\ref{sec:schwinger-lattice} we review our lattice setup, and in
Sec.~\ref{sec:num} we present our numerical results in the isospin symmetric
limit as well as for nondegenerate fermion masses. We present our conclusions in
Sec.~\ref{sec:conclu}.


\section{The $N_f=2$ Schwinger model}
\label{sec:SM}

The Lagrangian of the $N_{f}=2$  Schwinger model is given by\footnote{We assume
a vanishing $\theta$ vacuum, $\theta=0$, throughout the paper.}
\begin{equation}
    {\mathcal L} = -{1 \over 4} F_{\mu\nu} F^{\mu\nu} + \sum_{i=1,2} \bar{\psi}_i \left(i\not\!\partial - g \not\!\!A-m_i\right)\psi_i .
\end{equation}
where $g$ is the gauge coupling
and $F_{\mu\nu} = \partial_{\mu}A_{\nu} - \partial_{\nu}A_{\mu}$.
In the limit of massless fermions, it can be solved by
bosonization~\cite{Coleman:1976uz,Belvedere:1978fj} and by path integral
methods~\cite{Gattringer:1993ec,Joos:1994vy}. The bosonized theory depends on
two independent bosonic fields, $\eta$ and $\varphi$,
\begin{align}
     \label{eq:intro:bosonized-2f-schwinger}
  \mathcal{L} =& \; {1\over 2} \partial_{\mu} \eta \partial^\mu \eta - \frac{1}{2} \mu^2 \eta^{2} + \frac{1}{2}\partial_{\mu}\varphi\partial^{\mu}\varphi \nonumber \\
  &+c m^2 \cos \left( \sqrt{2\pi}~\eta \right) \cos (\sqrt{2\pi}~\varphi),
\end{align}
where $c = e^{\gamma} / 2\pi$ with $\gamma$ the Euler
constant, and $2 m^2 = m_1^2+ m_2^2$. The connection with the original theory is
given by
\begin{align}
     i \bar{\psi}_i \gamma^\mu \psi_i \equiv{1 \over \sqrt{\pi}} \epsilon_{\mu\nu} \partial_\nu \phi_i,\;\;\; m_i \bar{\psi}_i \psi_i \equiv - c m^2_i \cos\sqrt{4\pi} \phi_i,
\end{align}
and
\begin{align}
     \eta = {1\over\sqrt{2}} (\phi_1+\phi_2), \;\; \varphi ={1\over\sqrt{2}}( \phi_1-\phi_2).
\end{align}
While the $\mathrm{U}(1)_A$ symmetry is broken by the anomaly, the theory has a
nonanomalous $\mathrm{U}(1)_{V} \times \mathrm{SU}(2)_L\times \mathrm{SU}(2)_R$
symmetry that is not broken spontaneously, according to the
Mermin--Wagner--Coleman theorem~\cite{Coleman:1973ci}. This global symmetry is
however not transparent in this bosonic formulation.

In the limit $m_i\to0$, the $\eta$ field, which is an isospin singlet, is massive and
analogous to the $\eta'$ in QCD. Its mass is twice as large as in
the $N_{f}=1$ case~\cite{Gattringer:1993ec},
\begin{align}
  \label{eq:eta-prime-mass-nf2}
  \left.M_{\eta'}^2\right|_{m_i =0} = \mu^2 = {2 g^2 \over \pi}.
\end{align}
The second boson, $\varphi$, is massless in the same limit. The correlation
functions of the scalar and pseudoscalar currents have been computed
analytically~\cite{Joos:1994vy,Dilger:1995pi,Georgi:2020jik} in this limit, and
their behavior at large distances can be written as
\begin{align}
  \langle P^{a}(x) P^{b}(0) \rangle \sim \delta_{ab} \frac{1}{\left| x \right|},
  \label{eq:pp}
\end{align}
where $\Psi=(\psi_1,\psi_2)$ and  $P^{a} = i\bar{\Psi}\sigma^{a} \gamma_{5} \Psi$ with $\sigma^{a}$
for $a=1,2,3$ the Pauli matrices. This correlator does not behave like the
propagator of a massless pseudoscalar meson: Feynman's propagator in two
dimensions reads
\begin{align}
  \Delta_{F}(x) = \frac{i}{2\pi} K_{0}[m\sqrt{x^2}],
\end{align}
which in the massless limit becomes
\begin{align}
\lim_{m\to0}\Delta_{F}(x) =  -\frac{i}{4\pi} \log(x^2).
\end{align}
This indicates that the massless asymptotic states in this theory should rather
be described as unparticles~\cite{Georgi:2007ek,Georgi:2020jik}. Moreover, the
scaling of the pseudoscalar and scalar correlators in Eq.~(\ref{eq:pp})
indicates that the scaling dimension of the pseudoscalar and scalar densities is $d=1/2$.

The theory becomes more interesting when fermion masses are small, but nonzero.
As long as the strong coupling limit is considered, $m\ll g$, the massive $\eta$
field can be integrated out and the low-energy effective theory can be
represented by a sine-Gordon model,
\begin{align}
    \mathcal{L} =   \frac{1}{2}\partial_{\mu}\varphi\partial^{\mu}\varphi +c m^2 \cos (\sqrt{2\pi}~\varphi).
\end{align}

The mass gap of this model has been studied using the WKB
approximation~\cite{Dashen:1974ci},
\begin{align}
  \label{eq:intro:wkb-mass-scaling}
M_{\pi}^{{\rm WKB}} = {3\over \pi} M^{\rm cl} \approx 2.07 m^{2 / 3} g^{1 / 3}.
\end{align}
Another expression for the soliton mass can be derived from semiclassical
methods in the limit of large masses~\cite{Gattringer:1999gt},
\begin{align}
  \label{eq:intro:gattringer-scaling-mpi}
M_{\pi}^{{\rm cl}} = e^{2 \gamma / 3} \frac{2^{5 / 6}}{\pi^{ 1 / 6}} m^{2 / 3} g ^{1 / 3} \approx 2.1633 \;  m^{2 / 3} g^{1 / 3}.
\end{align}
Lastly, the exact $S$-matrix of the sine-Gordon theory has been computed
analytically~\cite{Zamolodchikov:1995xk}. The spectrum of the theory contains
four bound states: three of them, corresponding to the soliton, antisoliton and
breather modes, form a degenerate isospin triplet of mass $M_\pi$, while an additional soliton-antisoliton bound state below threshold is an
isospin singlet of mass~\cite{Coleman:1976uz}
\begin{align}
M_{\sigma} = \sqrt{3} M_\pi.
\end{align}
The exact result for the mass gap in the sine-Gordon model
is~\cite{Smilga:1996pi}
 \begin{align}
 \label{eq:intro:smilga-scaling-mpi}
   M_{\pi}^{{\rm SG}} =&\; m^{2 / 3}g^{1 / 3} 2^{5 / 6} e^{\gamma / 3} \left( \frac{\Gamma(3 / 4)}{\pi \Gamma(1 / 4)} \right)^{2 / 3} \frac{\Gamma(1 / 6)}{\Gamma(2 / 3)} \nonumber \\
   \approx&\; 2.008 \; m^{2 / 3} g^{1 / 3}.
\end{align}
The fermion condensate $\Sigma \equiv - \langle \bar{\psi}_{i} \psi_{i} \rangle$ is no longer vanishing in the presence of fermion masses
\cite{Smilga:1992hx} and the exact sine-Gordon prediction for this quantity
is~\cite{Smilga:1996pi}
\begin{align}
  \Sigma^{{\rm SG}} =&\; m^{1 / 3} g^{2 / 3} \frac{2^{2 / 3} e^{2 \gamma / 3}}{3 \sqrt{3} \pi^{4 / 3}} \left( \frac{\Gamma(3 / 4)}{\Gamma(1 / 4)} \right)^{4 / 3}  \frac{\Gamma(1 / 6)^2}{\Gamma(2  /3)^2} \nonumber \\
  \approx&\; 0.388 \; m^{1 / 3} g^{2 / 3}.
\label{eq:sigma}
\end{align}

The fermion mass scalings of the hadron masses, the condensate and
the decay constant (see next section) are consistent with the relations derived
in Refs.~\cite{Luty:2008vs,DelDebbio:2010ze,DelDebbio:2010jy} with the
appropriate modifications for two dimensions.\footnote{In the notation of
Refs.~\cite{DelDebbio:2010ze,DelDebbio:2010jy}, for this model we
find $\gamma^*=1/2$ and $y_m=3/2$. The scaling of the condensate in 2D is
$\langle \bar{q} q\rangle \propto m^{1-\gamma^*\over y_m}$, while that of the
matrix elements, $G_{\mathcal O} = \langle 0| {\mathcal O} |M\rangle$,  is
$G_{\mathcal O} \propto m^{\Delta_{\mathcal O}/y_m}$---compare with Eq.~(51) of
Ref.~\cite{DelDebbio:2010ze}---where $M$ is a meson state and
$\Delta_{\mathcal O}$ is the dimension of the operator ${\mathcal O}$. }
We note however that the sine-Gordon limit of the theory has been challenged in Refs.~\cite{Azcoiti:2019moz,Azcoiti:2021gst}.
One of the goals of our study is to test the validity of the different
approximations against our numerical lattice simulations. 

\section{Chiral Ward Identities}
\label{sec:WI}

Ward identities (WIs) are exact relations imposed by symmetries on correlation
functions.  We revisit the derivation of the Gell-Mann–Oakes–Renner (GMOR)
relation~\cite{Gell-Mann:1968hlm} which relates the axial current matrix
element $F_\pi$ with the fermion condensate and the triplet pseudoscalar meson
mass in the chiral limit,
\begin{align}
\lim_{m\rightarrow 0} {M_{\pi}^{2} \over 2 m}=   {\Sigma\over F_{\pi}^{2}}.
\end{align}
Given the absence of spontaneous symmetry breaking in this
theory, $\lim_{m\rightarrow 0} \Sigma=0$, the question is to what extent the relation holds for nonvanishing masses,
\begin{align}
  \label{eq:GMOR}
F_{\pi}^{2} M_{\pi}^{2} = 2 m \Sigma(m).
\end{align}

\subsection{Derivation of the GMOR relation}

We first recall the standard derivation of the GMOR relation~\cite{Luscher:1998pe}. The starting point
is the nonsinglet chiral WI:
\begin{align}
  \label{eq:WI-relation}
\partial^x_\mu \langle     A^a_\mu(x) P^b(y) \rangle= 2 m \langle P^a(x) P^b(y) \rangle  - {\delta_{ab}\over N_f} \delta(x-y) \langle S(y)\rangle,
\end{align}
where $A^a_\mu = \bar{\Psi} \gamma^\mu \sigma^a \Psi$ and $S= \bar{\Psi} \Psi$.

  We can
in all generality write the correlation function as
\begin{align}
\langle A^a_\mu(x) P^b(0) \rangle =\delta^{ab} x^\mu f(x^2),
\label{eq:APk}
\end{align}
for some arbitrary function $f$. Substituting in the WI with $y=0$ and $m=0$,
\begin{align} 2 f(x^2) + 2 x^2 f'(x^2)= 0,
    \end{align}
for $x\neq 0$. The solution of this equation is just
    \begin{align}
f(x^2) = {k \over x^2},
        \end{align}
where $k$ is a constant to be determined.
We can also consider the spectral decomposition of the same two-point
function: assuming dominance of the pion pole,
\begin{align}
  \langle A^a_\mu(x) P^b(0) \rangle =i \delta^{ab} F_\pi G_\pi \partial_\mu \Delta_\pi(x),
  \label{eq:APspec}
\end{align}
where $\Delta_\pi$ is the massless scalar propagator in two dimensions,
\begin{align}
 \Delta_\pi(x) = -{i \over 4 \pi} \log x^2,
\end{align}
and
\begin{align}
\langle 0| A_\mu^a |\pi(p)\rangle = i  F_\pi p_\mu , \;\;\;  \langle \pi(p) | P^a |0\rangle = G_\pi.
\end{align}
Matching Eqs.~(\ref{eq:APk}) and (\ref{eq:APspec}), we get
\begin{align}
{F_\pi G_\pi \over 2 \pi} = k.
\label{eq:k}
\end{align}
The WI also implies a relation between the matrix elements,
\begin{align}
\langle 0| \partial_\mu A_\mu^a(x) |\pi(p)\rangle = 2 m     \langle 0 | P^a(x) |\pi(p)\rangle,
\end{align}
or
\begin{align}
    F_\pi M_\pi^2 = 2 m G_\pi.
    \label{eq:Gpi}
\end{align}
Let us finally consider the integrated WI in the limit $m\rightarrow 0$ and
assume there is a nonvanishing condensate. In this case we would have
\begin{align}
\int d^2x\; \partial_\mu \langle     A^a_\mu(x) P^b(0) \rangle =&\;
\delta^{ab} \int_{R} d\sigma_\mu {k x^\mu\over x^2}  = 2\pi k \delta^{ab} \nonumber \\
  =&\; -\delta^{ab}{\langle S\rangle\over N_f} = \Sigma \delta^{ab},
\end{align}
where the integral on the right is a surface integral on a hypersphere of
radius $R$, with $d\sigma_{\mu}$ the infinitesimal area element on the surface
in the direction $\mu$. It follows
\begin{align}
  \label{eq:k-result}
k = \frac{\Sigma}{2\pi},
\end{align}
and substituting in Eqs.~(\ref{eq:k}) and (\ref{eq:Gpi}) the GMOR relation follows.

However, in our case $\Sigma$ vanishes in the chiral limit and a more careful
analysis is needed. In particular, we need to keep the term proportional to $m$
in Eq.~(\ref{eq:WI-relation}). Following an analogous derivation, one finds that,
at leading order in the fermion mass, the GMOR relation in Eq.~(\ref{eq:GMOR})
still holds at leading order in an expansion in $m$,  using the mass dependence of the chiral
condensate in Eq.~(\ref{eq:sigma}).

\subsection{Axial current and pseudoscalar matrix elements}

From the Ward identity one can also derive the first order scaling of $F_{\pi}$
and $G_{\pi}$ with the quark mass. Assuming that the KL decomposition is
saturated by the pole of the pion,
\begin{align}
\langle A^{a\mu} (x) P^{b}(0) \rangle = \delta_{ab}\frac{M_{\pi}}{2\pi} F_{\pi} G_{\pi} K_{0}^{\prime}(M_{\pi}\sqrt{x^2}) \frac{x^{\mu}}{\sqrt{x^2}},
\end{align}
where the prime represents derivative with respect to the argument. Knowing that in the chiral limit
\begin{align}
\lim_{M_{\pi}\to0} K'_{0}[M_{\pi}\sqrt{x^2}] = \frac{1}{M_{\pi}}\frac{1}{\sqrt{x^2}},
\end{align}
and comparing with Eq.~(\ref{eq:APk}), one
derives $k\sim G_{\pi} F_{\pi}$. Finally, since from
Eq.~(\ref{eq:k-result}) we know that $k\sim \Sigma \sim m^{1 / 3}$, we can use
Eq.~(\ref{eq:GMOR}) and the mass scaling of the pion mass to derive
\begin{align}
  \label{eq:intro:fpi-gpi-scaling}
F_{\pi} \sim m^{0}, \quad G_{\pi} \sim m^{1 / 3}.
\end{align}
From this we can see another striking difference with respect to QCD: the
overlap of the pseudoscalar density and the one-pion state $G_{\pi}$ vanishes
in the chiral limit, further indicating that pions ``dissolve'' into unparticles in this
limit~\cite{Georgi:2007ek}.

\vspace{0.5cm}

Additionally, from the GMOR relation we can get a prediction for $F_\pi$, which is
dimensionless in two dimensions. The analytical results of  $M^{\rm SG}_\pi$
and $\Sigma^{\rm SG}$ in Eqs.~(\ref{eq:intro:smilga-scaling-mpi}) and
(\ref{eq:sigma}), combined with the GMOR relation in Eq.~(\ref{eq:GMOR}), give
\begin{align}
    (F_\pi^{\text{SG}})^2 = {2 m\Sigma^{\text{SG}}\over (M_\pi^{\text{SG}})^2} = {1\over 3 \sqrt{3}}.
\end{align}
This can be compared with the prediction of $F_{\eta'}$ from the Witten--Veneziano relation. The
Witten--Veneziano relation is exact in the chiral limit of this model. The
topological charge density correlator can be computed analytically at nonzero
momentum and it is saturated exactly by the $\eta'$
pole and reads~\cite{Seiler:1987ig,Gattringer:1993ec,Giusti:2001xh}
\begin{align}
  \label{eq:quenched-WV}
\lim_{m \rightarrow 0} {F_{\eta'}^2 M_{\eta'}^2\over 2 N_f} = \chi_{\rm top}^{\text{quenched}},
\end{align}
where
$F_{\eta'} \equiv M_{\eta'}^{-2}\langle 0| \partial_\mu A_\mu |\eta'\rangle$,
and the topological susceptibility in
the pure gauge theory is
\begin{align}
\chi_{\text{top}}^{\text{quenched}} = \frac{g^2}{4\pi^2}.
\end{align}
From the two previous equations and identifying $\mu^2=M_{\eta'}^2$ in
Eq.~(\ref{eq:eta-prime-mass-nf2}) it follows
\begin{align} (F_{\eta'}^{\text{WV}})^2 = {1 \over 2\pi}.
\label{eq:fetap}
\end{align}
In  the limit of QCD with large number of
colors, $N_{c} \to \infty$, it can be shown that $F_{\eta'}=F_{\pi}$. We note that the Witten--Veneziano
relation in Eq.~(\ref{eq:quenched-WV}) is inconsistent
with $F_{\eta'}=F_{\pi}$ in this case. On the other hand, recent simulations seem to indicate
that $F_{\pi}$ is close to Eq.~(\ref{eq:fetap})~\cite{Castellanos:2023rdw}.

\section{Low-energy Effective Theory and Isospin Breaking}
\label{sec:EFT}

The bosonized Lagrangian in Eq.~(\ref{eq:intro:bosonized-2f-schwinger}) does not
provide a transparent representation of the isospin multiplets of the theory.
The degeneracy of the soliton, antisoliton and breather mode can be guessed from
the global symmetry of the theory, but it looks miraculous from the solution
of the sine-Gordon theory.

In the strong coupling limit, $m\ll g$, there is a clear separation of scales
since $M_\pi \ll \mu$. This suggests that a low-energy effective field theory (EFT) describing
only the light degrees of freedom can be constructed. The EFT should include
both the pions and the scalar singlet since the ratio of both masses is
just $\sqrt{3}$ and should ideally make the global flavor symmetry explicit.
In Ref.~\cite{Delphenich:1997ex}, the interesting observation was made that a
linear $\sigma$ model, together with the assumption that the quark condensate
must vanish in the chiral limit, predicts the ratio
\begin{align}
M_{\sigma} = \sqrt{3} M_{\pi}.
\label{eq:sqrt3}
\end{align}
A similar relation was found in \cite{PhysRevD.2.685} in the context of a chiral
EFT with spontaneous breaking of chiral and conformal symmetries, which
includes a dilaton.  However, in the proposal of Ref.~\cite{Delphenich:1997ex}
the scaling of the pion mass with the fermion mass is not properly reproduced.

Inspired by this, we consider a nonlinear $\sigma$ model, including a dilaton
field,  and show that it predicts the correct scaling of the pion with the quark
mass, as well as the ratio of masses in Eq.~(\ref{eq:sqrt3}). Furthermore, if we
also include the pseudoscalar singlet, the $\eta'$, as dictated to reproduce
the $U(1)_A$ anomaly, a prediction for the isospin breaking corrections in the
meson spectrum can be obtained.

We use a nonlinear parametrization of the pseudoscalar meson bilinears
including the scalar and pseudoscalar singlet mesons,
\begin{align}
U = e^{\sigma+ i \eta'+i {\pi^{a}} { \sigma^{a}}}.
\end{align}
Under $\mathrm{U}_L(2) \times \mathrm{U}_R(2)$ chiral rotations the field
transforms as
\begin{align}
U \rightarrow g_L U g^\dagger_R.
\end{align}
Under a scale transformation $x\rightarrow e^\lambda x$,
\begin{align}
    \sigma(x)\rightarrow \sigma(e^\lambda x) - {\rm d} \lambda,
\end{align}
where ${\rm d}$ is the scaling dimension of the scalar density operator, which
is ${\rm d}=1/2$ in the chiral limit as discussed, in Sec.~\ref{sec:SM}.

The most general Lagrangian which satisfies the chiral symmetry and scale
invariance is
\begin{align}
  \label{eq:schwinger:chiral-lagrangian}
{\mathcal L} = {1 \over 4} {\rm Tr}[L_\mu^\dagger L_\mu] - V[U],
\end{align}
where $L_\mu\equiv U^{-1}\partial_\mu U$ and
\begin{align}
V[U]=V_s[U]+V_m[U]+V_a[U].
\end{align}
Here, $V_s[U]$ is symmetric under the
nonanomalous $\mathrm{SU}(2)_L \times \mathrm{SU}(2)_R \times \mathrm{U}(1)_V$ and is
scale invariant,
\begin{align}
  V_s[U] =  a {\rm Tr}[U^\dagger U]^2+ b {\rm Tr}[(U^\dagger U)^2],
\end{align}
where $a$ and $b$ are low-energy couplings, unconstrained by symmetries.
Note that only terms with four powers of $U$ are scale invariant.
On the other hand, $V_m[U]$ is the mass term, which breaks chiral symmetry and
scale invariance. The mass term in the underlying theory
is $\bar{\psi}_R M \psi_L +\text{H.c.}$, which becomes chirally symmetric if we
take $M$ to be a spurion that transforms as $M\rightarrow g_R M g_L^\dagger$
and also scale invariant if $M$ scales as $M\rightarrow e^{3\lambda/2} M$.
At leading order in $M$, the only
term that is symmetric is then
\begin{align}
V_m[U] =- d {\rm Tr}[M U + U^\dagger M^\dagger].
\end{align}

Finally $V_a[U]$ implements the anomalous $\mathrm{U}(1)_A$ Ward identity~\cite{DiVecchia:1980yfw} in the effective theory,
\begin{align}
  V_a[U] = -{c\over 2}  (\log[\det U] -\log[\det U^\dagger])^2.
\end{align}
It is not scale invariant, because it involves the heavy sector of the theory,
i.e. the $\eta'$.

Strictly speaking, the heavy sector should not be part of the low-energy effective
theory. However, it is necessary to mediate isospin corrections, as we will see.
In the large $N_c$ limit of QCD, there is a justification to include the $\eta'$
in chiral perturbation theory because the $\eta'$ mass can be made small for
large enough $N_c$. Although the situation here is different, we expect that the
effect of including the anomaly term is equivalent to including higher-dimensional
operators suppressed by the heavy scale $m_{\eta'}$.

Considering the isospin symmetric limit, that is $M={\rm Diag}(m,m)$, the
minimization of the potential leads to a minimum at
\begin{align}
\langle \sigma\rangle = {1\over 3}  \log{\left({ d m\over 8 a + 4 b}\right)}.
\end{align}
Expanding the potential around this vacuum up to quadratic order, we find
the $\pi$, $\eta'$ and $\sigma$ masses to be
\begin{align}
M_\pi^2 =&  \left( {2^{8}d ^{4} \over a+{b\over 2}} \right)^{1 / 3}m^{4 / 3},\quad M^2_{\eta'}  = 16 c + M^2_{\bf \pi}, \nonumber \\
M^2_{\sigma} =&\; 3 M_\pi^2.
\end{align}
The chiral effective Lagrangian in Eq.~(\ref{eq:schwinger:chiral-lagrangian})
thus provides the expected quark mass scaling from the strong coupling limit of
the Abelian bosonization of the theory, while also predicting the correct
scalar-to-pseudoscalar mass ratio. It would be interesting to understand if the connection between the scalar and pseudoscalar masses might be generic in theories in 4D with conformal symmetry broken by mass terms.

\vspace{0.5cm}

One can add isospin breaking in this effective model by setting
\begin{align}
M ={\rm Diag}\left(m-{\Delta\over 2},m+{\Delta\over 2}\right)= m I_2 -{\Delta\over 2} \sigma_3,
\end{align}
where $I_2$ is the identity matrix in isospin space. While the vacuum expectation value does not change, the masses become
\begin{align}
M_{\pi^{\pm}}^2 =& \; \left( \frac{2^{8}d ^{4}}{a+ \frac{b}{2}} \right)^{1 / 3}m^{4 / 3}, \\
M_{\sigma}^{2} =& \; 3 M_{\pi^{\pm}}^{2}, \\
M_{\pi^{0}}^{2} =& \; M_{\pi^{\pm}}^{2} - \frac{1}{16c} \frac{2^{2 / 3} d^{8 /
                   3} \Delta^2 m^{2  /3}}{ \left( a + \frac{b}{2} \right)^{2 / 3}} \nonumber \\
  =&\; M_{\pi^{\pm}}^{^2} - \frac{1}{16c} \frac{M_{\pi^{\pm}}^{4}}{4} \left(
\frac{\Delta}{m} \right)^2, \label{eq:pion-mass-splitting} \\
M_{\eta'}^{2} =& \; 16c + 2 M_{\pi^{\pm}}^{2} - M_{\pi^{0}}^{2}.
\end{align}
The charged to neutral pion mass difference can then be written as
\begin{align}
  \label{eq:pion-mass-splitting-result}
M_{\pi^{\pm}}^{2} - M_{\pi^{0}}^{2} = \frac{1}{4}
\frac{M_{\pi^{\pm}}^{4}}{M_{\eta'}^{2}|_{m=0}} \left(  \frac{\Delta}{m} \right)^2,
\end{align}
with $M_{\eta'}^2|_{m=0} = 16c$, and thus confirms the findings in
Ref.~\cite{Delphenich:1997ex} that the charged to neutral pion splitting is
proportional to the square of the quark mass differences and suppressed in
the square of the $\eta'$ mass. Note that Eq.~(\ref{eq:pion-mass-splitting-result}) is a parameter-free prediction.

The isospin symmetry in the light spectrum is therefore accidental: it is a consequence of the fact that the only operator we can write down in the effective theory that breaks isospin symmetry identically vanishes, i.e. ${\rm Tr}[\sigma_3 (U+U^\dagger)] =0$ for $\eta'=0$.

Finally, the concept of {\it automatic fine-tuning} of Ref. \cite{Georgi:2020jik} due
to the exponentially small isospin breaking corrections is somewhat misleading:
if one considers the correlation function of two isospin breaking operators, it is exponentially suppressed in
the operator separation as $\exp(-M_{\eta'} x)$ \cite{Georgi:2020jik}; however, there are isospin breaking corrections to the spectrum that are just suppressed in inverse powers of the heavy scale.
The isospin breaking corrections are mediated by the pseudoscalar singlet meson,
the $\eta'$, which is heavy and decouples from the EFT. The  pseudoscalar meson
splitting corresponds, in the EFT without $\eta'$, to a higher-dimensional
operator induced by the integration of this heavy scale.  The first correction
appears at second order in $\Delta$ and is suppressed by the $\eta'$ propagator
at low momentum by $M_{\eta'}^{-2}$, as seen in Eq.~(\ref{eq:pion-mass-splitting-result}).

\section{Schwinger model on the lattice}
\label{sec:schwinger-lattice}

The lattice formulation of the theory relies on the discretized Euclidean partition function
\begin{align}
Z = \int \mathcal{D}U\mathcal{D}\bar{\psi}\mathcal{D}\psi\; e^{-S_{G}[U]-S_{F}[U, \psi,\bar{\psi}]},
\end{align}
with the integration measure
\begin{align}
\mathcal{D}U = \prod_{x,\mu}^{} dU_{x,\mu}, \quad \mathcal{D}\psi = \prod_{x,i}^{}d\psi_{x,i}, \quad \mathcal{D}\bar{\psi} = \prod_{x,i}^{}d\bar{\psi}_{x,i},
\end{align}
and $U_{x,\mu}\in\mathrm{U}(1)$ being the gauge link living on the lattice
edge connecting the points $x$ and $x+\hat{\mu}$ of the two-dimensional lattice grid, with $\hat{\mu}$ a unit
vector in the $\mu$th direction. We consider a square lattice
of size $L \times L$ with periodic boundary conditions.

We use the Wilson discretization of the gauge action, which reads
\begin{align}
  \label{eq:lat-meth:U1-gauge-action}
S_{G}[U] = -\beta \sum_{x \in \Lambda}^{} {\rm Re}[U_{p}(x)],
\end{align}
where $\beta = 1/g^2$  and $U_{p}(x)$ is the $1\times1$ Wilson loop at the
lattice point $x$,
\begin{align}
U_{p}(x) &= U_{x,0} U_{x+\hat{0}, 1} U_{x+\hat{1},0}^{\dagger} U_{x,1}^{\dagger}.
\end{align}
Note that all dimensionful quantities are assumed in lattice units.
Particularly, $\beta$ is dimensionful and it scales with the lattice
spacing $a$ as $\beta \sim a^{-2}$.

We also use the Wilson discretization of the fermion action,
\begin{align}
  \label{eq:latmeth:qcd-fermion-action}
	S_{F}[U, \psi,\bar{\psi} ] = \sum_{i}\sum_{x,y\in\Lambda}^{}\bar{\psi_i}(x) K_{i}(x,y) \psi_i(y),
\end{align}
where the Dirac operator for the flavor $f$ reads
\begin{align}
  K_{i}(x,y) =\;(m_{i} + 2)\delta_{xy}
  &-\frac{1}{2} \sum_{\mu} \left[ \left( 1-\gamma_{\mu} \right)U_{x,\mu}\delta_{y,x+\hat{\mu}} \right. \nonumber \\
  &\left.+(1+\gamma_{\mu})U^\dagger_{x-\hat{\mu},\mu}\delta_{y,x-\hat{\mu}} \right] .
\end{align}
The integration over the fermion fields can be done exactly
 and yields the product of
determinants $\prod_{i}^{}\det K_{i}$. For two degenerate flavors, the
determinant can be computed stochastically introducing a complex bosonic
field $\phi$,
\begin{align}
\det K \det K = \det K K^{\dagger} = \int \mathcal{D}\phi \; e^{-S_{\text{pf}}[U,\phi]},
\end{align}
with the pseudofermion action
\begin{align}
  \label{eq:lat-meth:U1-pseudofermion-action}
S_{\text{pf}}[U, \phi] = \sum_{x,y\in \Lambda}^{} \phi(x)^{\dagger}(KK^{\dagger})^{-1}_{x,y}\phi(y).
\end{align}
We simulate the theory using a modification of the hybrid Monte Carlo
(HMC) algorithm~\cite{Duane:1987de}, refered to as the winding HMC
algorithm~\cite{Albandea:2021lvl}, which has been shown to significantly improve
the sampling efficiency of the different topological sectors in this theory.

For the case of nondegenerate fermions, we use the rational HMC (RHMC)
algorithm~\cite{Horvth1999,Clark2004} with the pseudofermion action
\begin{align}
  \label{eq:lat-meth:U1-pseudofermion-action}
S_{\text{pf}}[U, \phi] = \sum_{i}\sum_{x,y\in \Lambda}^{} \phi_i(x)^{\dagger}\sqrt{(K_{i}K_{i}^{\dagger})^{-1}_{x,y}}\phi_i(y).
\end{align}

\section{Numerical Results}
\label{sec:num}

\subsection{Degenerate case}

\subsubsection{Pion mass dependence on the quark mass}
\label{eq:windings:mpi-dependence-m}

As we saw in Sec.~\ref{sec:SM}, in the degenerate two-flavor
Schwinger model the mass of the pion is expected to scale as
\begin{align}
  \label{eq:windings:expected-mpi-scaling}
M_{\pi} = A m_{R}^{2 / 3} g^{1 / 3},
\end{align}
where $m_{R}$ is the renormalized quark mass and $A$ is a proportionality
constant. This constant has been derived from a semiclassical approximation
in the strong coupling limit~\cite{Gattringer:1999gt}, the WKB
approximation for the sine-Gordon theory~\cite{Dashen:1974cj}, as well as
exactly in the latter~\cite{Smilga:1996pi}, leading to
\begin{align}
  \label{eq:factor-scaling-M-vs-m}
A^{\text{cl}} \approx 2.16, \quad A^{\text{WKB}} \approx 2.07, \quad A^{\text{SG}} \approx 2.008,
\end{align}
respectively.

\begin{figure}[!t]
	\centering
	\includegraphics[width=\linewidth,keepaspectratio,angle=0]{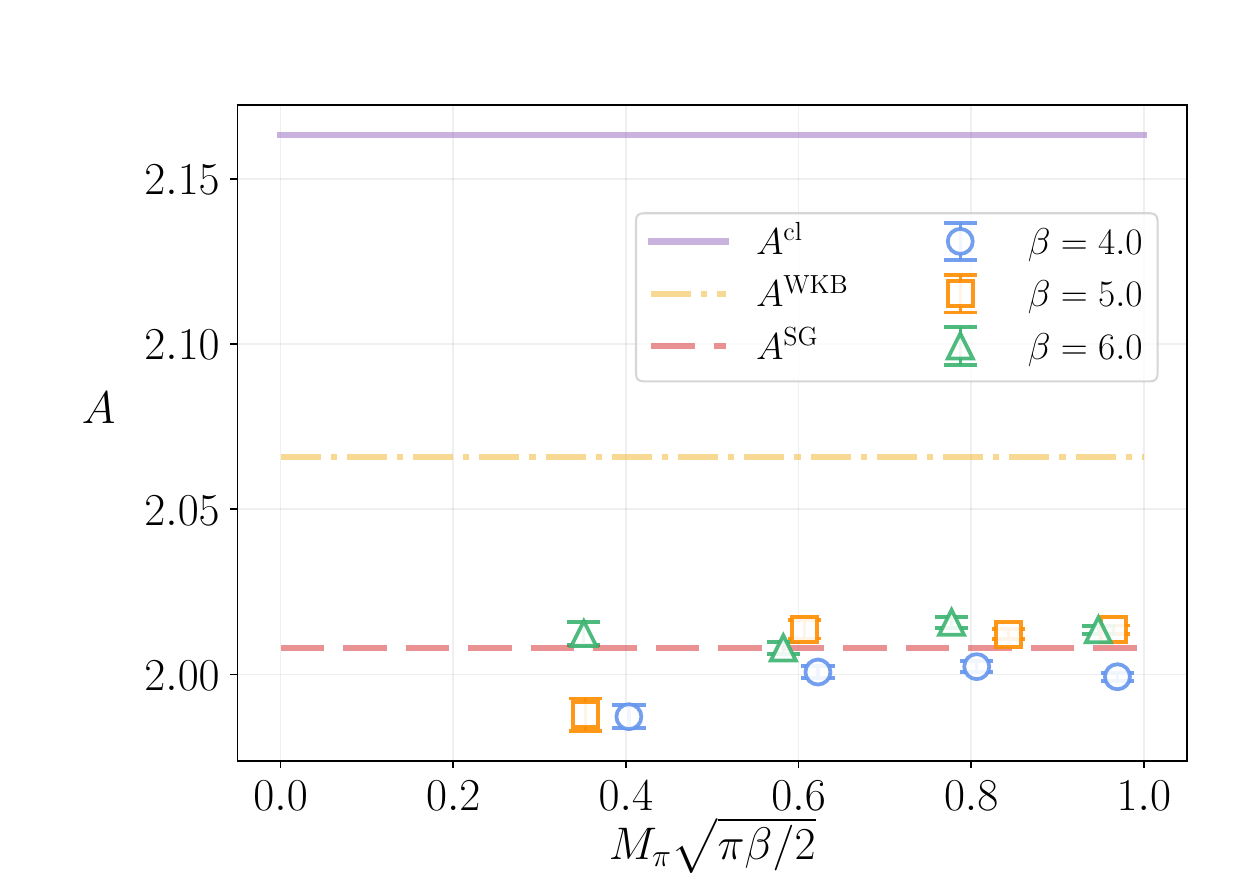}
	\caption{Proportionality factor in
    Eq.~(\ref{eq:windings:expected-mpi-scaling}) as a function
    of $M_{\pi} / \mu = M_{\pi} \sqrt{\pi \beta / 2}$. Blue circles, orange
    squares and green triangles correspond to simulations with $V=64 \times 64$
and $\beta=4,5$ and $6$, respectively. The solid, dash-dotted and dashed lines
correspond to
  the semiclassical, WKB and sine-Gordon results of
  Eq.~(\ref{eq:factor-scaling-M-vs-m}).}
	\label{fig:Mpivsm}
\end{figure}

The renormalized mass in Eq.~(\ref{eq:windings:expected-mpi-scaling}) is the
partially conserved axial current mass, defined from the nonsinglet axial Ward identity,
\begin{align}
\bra{0} \partial_{\mu} A_{\mu}^{a} \ket{\pi(p)} = 2 m_{R} \bra{0} P^{a} \ket{\pi(p)}.
\end{align}
We can obtain $m_{R}$ by inspecting the plateau of the ratio
\begin{align}
m_R = \frac{1}{2} \frac{\bra{0} [A_{\mu}^{a}(x+a)-A_{\mu}^{a}(x)]O(x)\ket{0}}{\bra{0} P^{a}(x) O(x) \ket{0}},
\end{align}
where $O$ is an interpolator coupling to the pion and for which we
choose $O(x)=P^{a}(x)$. With Wilson fermions, the renormalized mass is related to the bare mass by
\begin{align}
m_{R} = Z_{m}(m - m_{c}),
\end{align}
where $m$ is the bare quark mass and $m_{c}$ is the critical mass for
which $m_{R}$ vanishes.

We want to study the approach to the strong coupling limit, $M_{\pi} \ll \mu$,
where $\mu$, defined in Eq.~(\ref{eq:eta-prime-mass-nf2}), is the mass of the pseudoscalar singlet in the
chiral limit. We have performed simulations at different values of $M_{\pi} / \mu$ in the
range $[ 0.3, 1.0 ]$, and the corresponding results are displayed
in Table~\ref{tab:mpi-vs-mr}. A similar study was carried out in
Ref.~\cite{Gattringer:1999gt} for a lattice volume $V=32\times 32$ and coupling
values $\beta=4,5,6$, finding a reasonably good agreement
with $A^{\text{cl}}$ for large values
of the mass. However, the statistical errors close to the chiral limit made the
agreement with $A^{\text{SG}}$ unclear for small masses.

\begin{figure*}[t!]
	\centering
	\includegraphics[width=0.49\linewidth,keepaspectratio,angle=0]{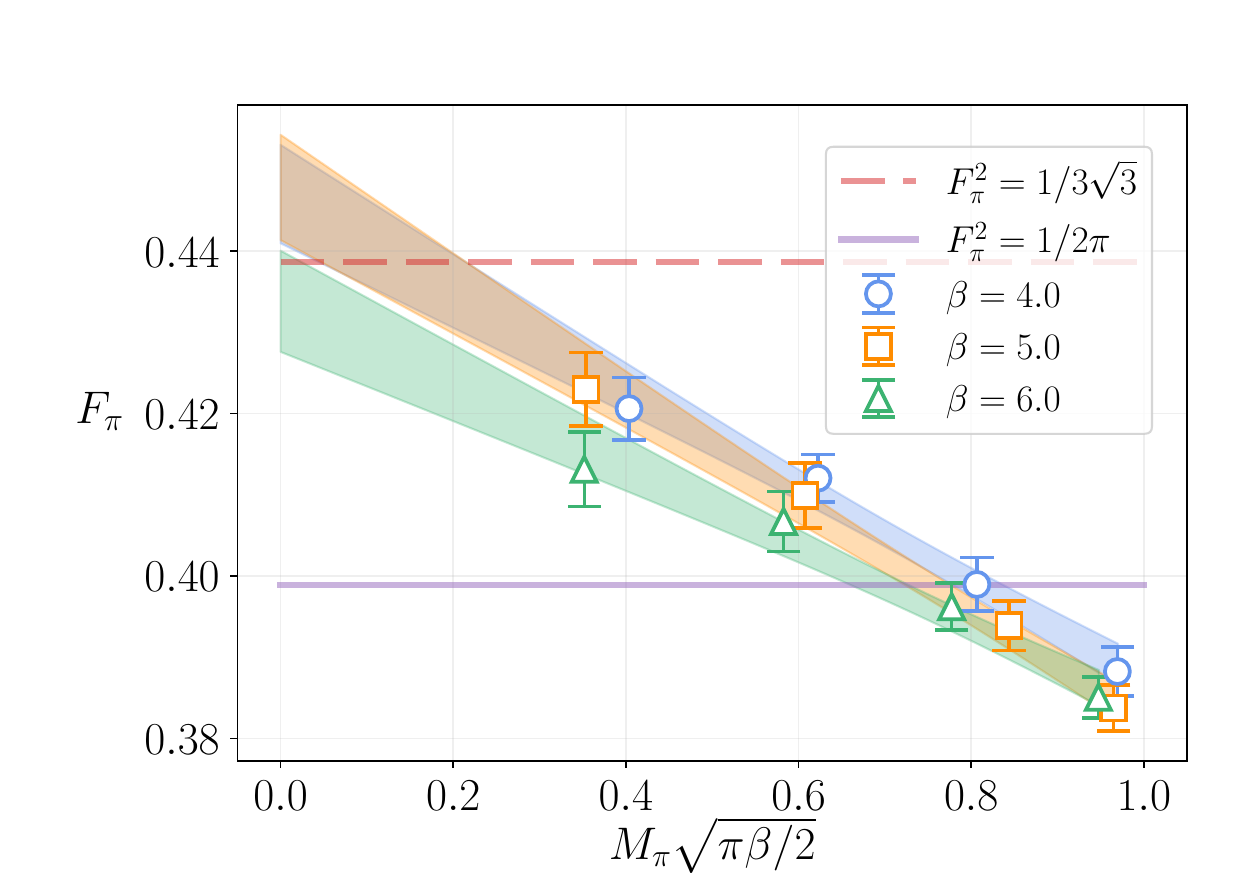}
	\includegraphics[width=0.49\linewidth,keepaspectratio,angle=0]{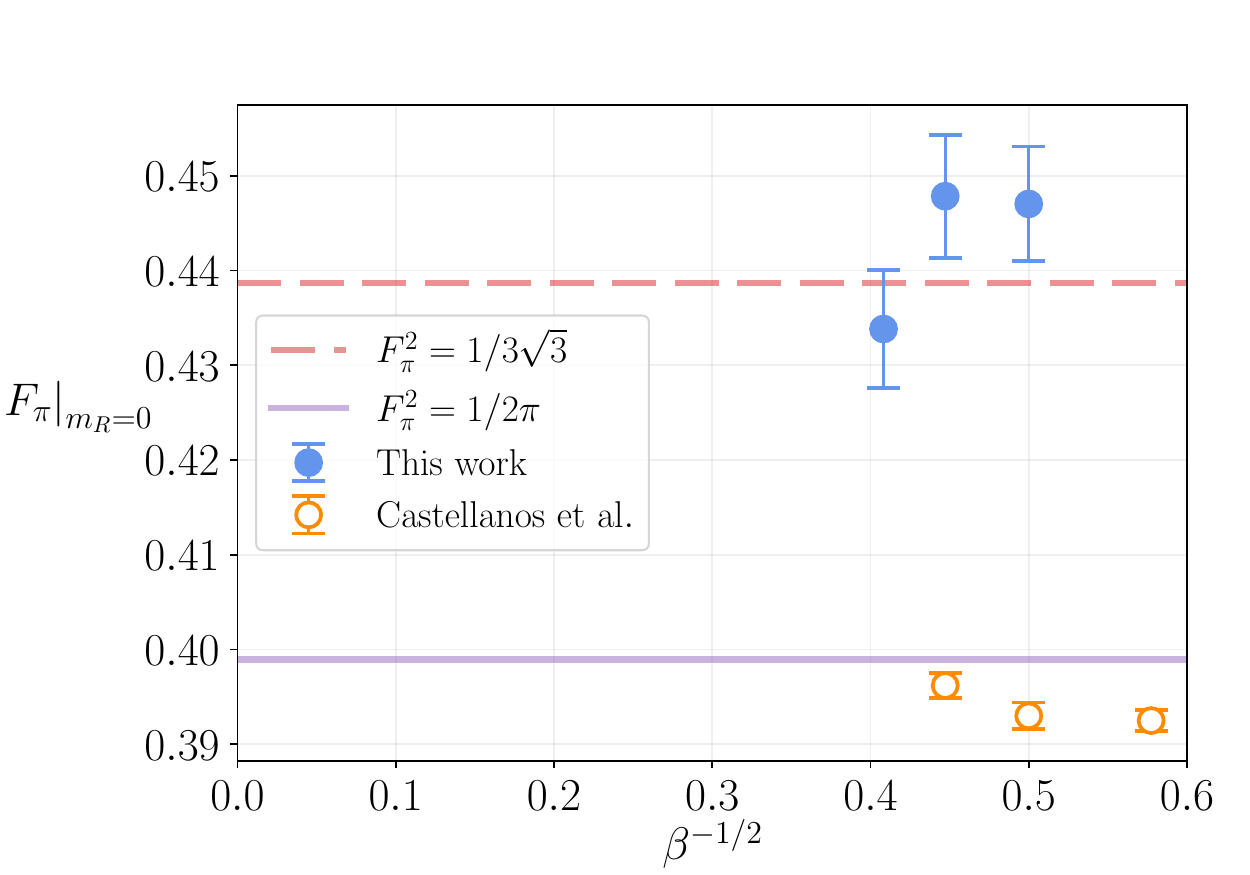}
	\caption{Left: pion decay constant as a function of $M_{\pi} / \mu$
    simulation at coupling values $\beta =\{ 4,5,6 \}$ (blue circles, orange
    squares, green triangles) and
different masses of the degenerate quark masses. The prediction coming from the
sine-Gordon theory, $F^{\text{SG}}$, and the Witten--Veneziano
relation, $F^{\text{WV}}$, are represented by a dashed and a solid line,
respectively. The bands correspond to a linear chiral limit extrapolation.
Right: values of $F_{\pi}$ extrapolated to the chiral limit, $m_{R}=0$, as a
function of $\beta^{- 1 / 2}$. The results from this work are displayed in full
blue circles, while in open orange circles we show the results from
Ref.~\cite{Castellanos:2023rdw}.}
	\label{fig:Fpi}
\end{figure*}

\begin{table}[!t]
  \centering
\begin{tabular}{c|c|c|c|c|c}
$\beta$ &  $m$ & $m_{R}$ & $M_{\pi}$ & $F_{\pi}$ & $G_{\pi}$\\
  \hline\hline
4.0 &$0.02$&  0.1203(10) & 0.38674(24) & 0.3882(30) & 0.35200(73)\\
 &   $-0.01$&  0.09107(88) & 0.32172(26) & 0.3990(33) & 0.32681(78)\\
 &   $-0.04$&  0.06183(75) & 0.24833(22) & 0.4121(29) & 0.29218(64)\\
 &   $-0.07$&  0.03260(65) & 0.16097(28) & 0.4206(38) & 0.24276(80)\\
  \hline
5.0 &$ 0.025$& 0.10575(87) & 0.34434(23) & 0.3838(28) & 0.31134(62)\\
 &   $ 0.005$&  0.08649(77) & 0.30095(23) & 0.3939(31) & 0.29446(62)\\
 &   $-0.03 $&  0.05278(62) & 0.21669(30) & 0.4099(40) & 0.25684(78)\\
 &   $-0.06 $&  0.02389(54) & 0.12612(31) & 0.4230(45) & 0.20161(85)\\
  \hline
6.0 &$ 0.025$& 0.09390(79) & 0.30857(20) & 0.3851(25) & 0.28112(49)\\
 &   $ 0.0  $& 0.06967(66) & 0.25318(23) & 0.3962(29) & 0.25902(54)\\
 &   $-0.025$& 0.04545(56) & 0.18969(28) & 0.4067(37) & 0.22747(70)\\
 &   $-0.05 $& 0.02122(49) & 0.11442(31) & 0.4131(46) & 0.18329(84)\\
 \hline\hline
\end{tabular}
    \caption{Values of, $m$, $m_{R}$, $M_{\pi}$, $F_{\pi}$ and $G_{\pi}$ for the different
      simulations at $\beta=4,5,6$ and $V=64\times64$.}
  \label{tab:mpi-vs-mr}
\end{table}

\begin{table}[!t]
  \centering
    \begin{tabular}{c|c|c}
      $\beta$ & $Z_{m}$ & $m_{c}$ \\
      \hline
      $4.0$ &  $0.9745(67)$  & $-0.10345(63)$ \\
      $5.0$ &  $0.9630(63)$  & $-0.08481(54)$\\
      $6.0$ &  $0.9690(66)$  & $-0.07190(48)$\\
      \hline
    \end{tabular}
    \caption{Values of $Z_{m}$ and critical mass $m_{c}$ for the different
      simulations at $\beta=4,5,6$.}
  \label{tab:windings:beta-zm-mc}
\end{table}

While we study the same values of the coupling, we perform simulations at a
lattice volume $V=64\times64$ to reduce finite size effects. For each $\beta$,
we obtain both $Z_{m}$ and $m_{c}$ from a conventional fit to a straight line,
and the results are shown in Table~\ref{tab:windings:beta-zm-mc}. In
Fig.~\ref{fig:Mpivsm} we show the proportionality constant in
Eq.~(\ref{eq:windings:expected-mpi-scaling}) as a function of $M_{\pi} / \mu$. Although the numerical difference between the different
approximations is small, there is enough statistical significance to conclude
that in these range of masses the results are indeed compatible with
the exact solution of the sine-Gordon theory, $A^{\text{SG}}$, and differ
significantly with the other approximations. In particular, at the finest lattice spacing
the value perfectly agrees with the exact result of the strong coupling limit
approximation for the smallest values of $M_{\pi} / \mu$, as expected.

\subsubsection{Pion decay constant and matrix element}

\begin{figure*}[t!]
	\centering
	\includegraphics[width=0.49\linewidth,keepaspectratio,angle=0]{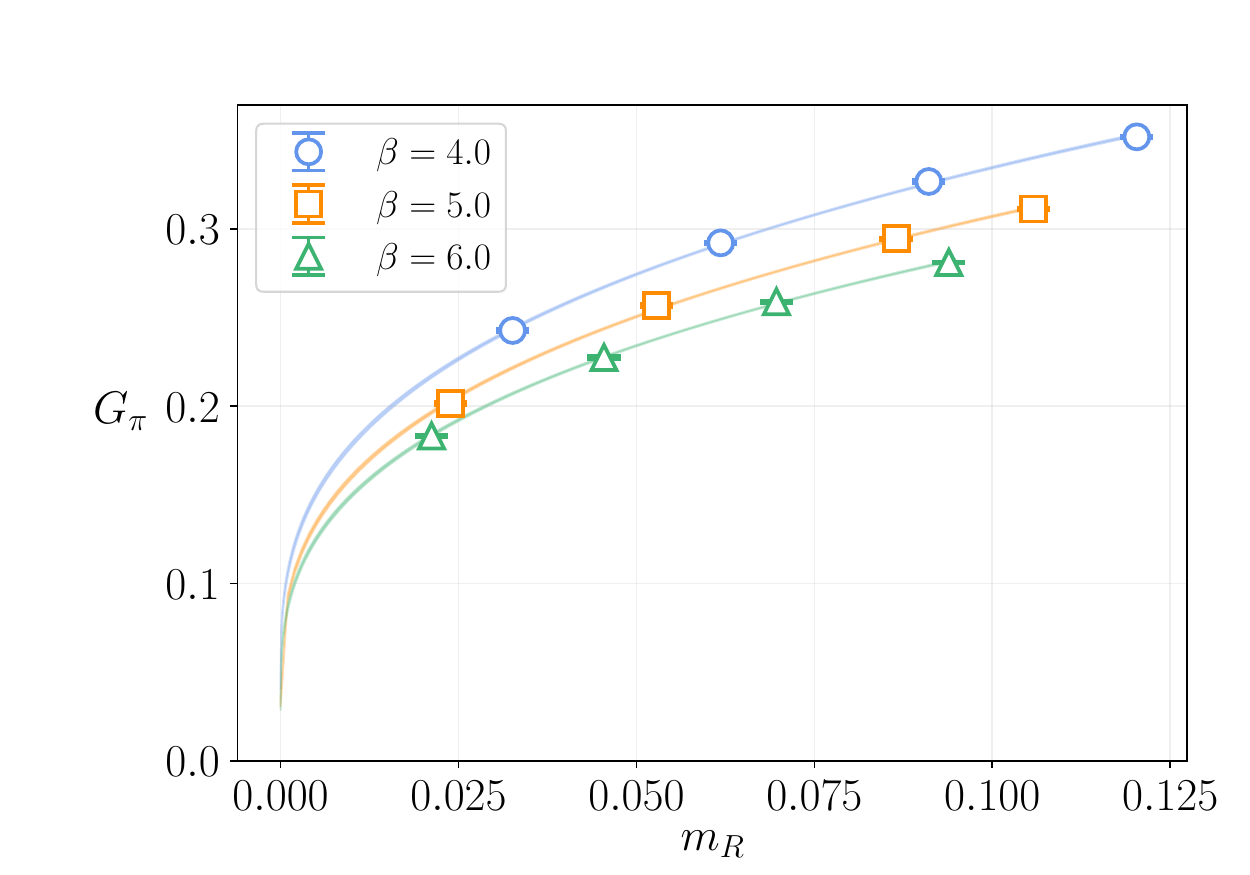}
	\includegraphics[width=0.49\linewidth,keepaspectratio,angle=0]{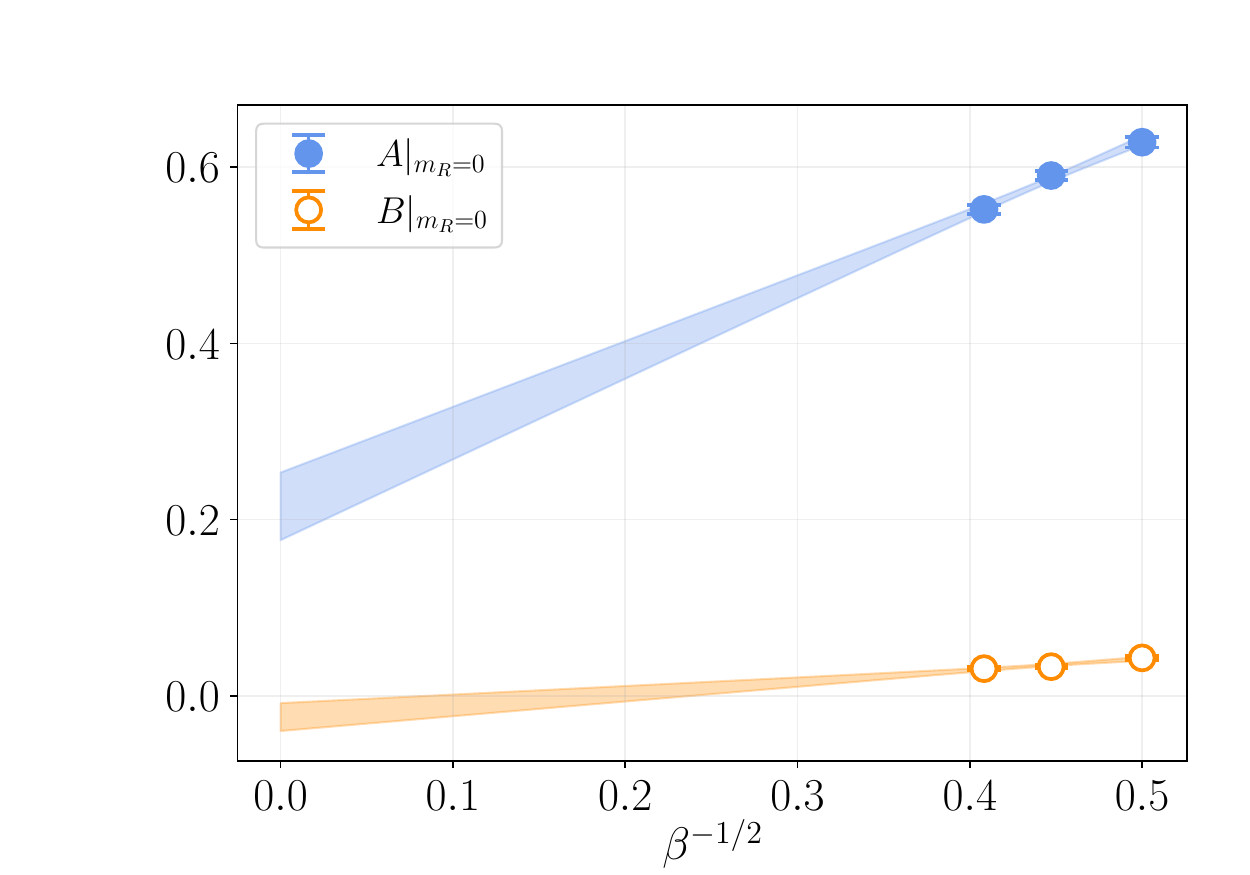}
	\caption{Left: pion matrix element $G_{\pi}$ as a function of the
    renormalized mass $m_{R}$ from
    simulation at coupling values $\beta =\{ 4,5,6 \}$ (blue circles, orange
    squares, green triangles) and
different values of the degenerate quark masses. The bands correspond to fits to
the functional form $G_{\pi}(m_{R}) = A m_{R}^{1 / 3} + B$. Right: fitting
parameters $A$ (full blue circles) and $B$ (open orange circles) extrapolated
to the chiral limit $m_{R}=0$ as a function of $\beta^{-1 / 2}$. The bands
correspond to a linear continuum extrapolation.}
	\label{fig:Gpi}
\end{figure*}

We saw in Sec.~\ref{sec:WI} two different predictions for the pion decay
constant,
\begin{align}
F_{\pi}^{\text{SG}} = \frac{1}{\sqrt{3 \sqrt{3}}}, \quad  F_{\pi}^{\text{WV}} = \frac{1}{\sqrt{2\pi}}.
\end{align}
While $F_{\pi}^{\text{SG}}$ is obtained from the exact solution of the sine-Gordon
model through the GMOR relation and is therefore expected to hold in the chiral
limit, $F_{\pi}^{\text{WV}}$ is expected to hold
if $F_{\eta'} = F_{\pi}$ from the Witten--Veneziano relation and from semiclassical approximations. This relation is
true in QCD at large $N_{c}$, but there is no reason why it should hold in the Schwinger model.

The pion matrix element can be obtained from lattice simulations by fitting the
pseudoscalar-pseudoscalar current to the functional form
\begin{align}
\sum_{x_1=0}^{L-1} \langle  P^{a}(x) P^{a}(0) \rangle = G_{\pi}^{2} \frac{\cosh \left[ M_{\pi} \left(  x_0 - \frac{L}{2} \right) \right]}{2 M_{\pi} \sinh \left[ M_{\pi} \frac{L}{2} \right]},
\end{align}
for sufficiently large separations $x_{0}$, where $x\equiv(x_0, x_1)$. For the
pion decay constant we additionally need the axial-pseudoscalar current,
\begin{align}
\sum_{x_1=0}^{L-1}\langle  A^{0a}(x) P^{a}(0) \rangle = \sqrt{2} F_{\pi} M_{\pi} G_{\pi} \frac{\sinh \left[ M_{\pi} \left( x_0 - \frac{L}{2} \right) \right]}{2 M_{\pi} \sinh \left[ M_{\pi} \frac{L}{2} \right]}.
\end{align}

We have computed both quantities for the same simulation parameters as those
reported in Sec.~\ref{eq:windings:mpi-dependence-m}. In
Fig.~\ref{fig:Fpi} (left) we show the results for the pion decay
constant $F_{\pi}$ as a function $M_{\pi} / \mu$, and find that the chiral limit
extrapolation of the simulations at coupling values $\beta=\{ 4,5,6 \}$
seem to support the sine-Gordon
prediction, $F_{\pi}^{\text{SG}}$, corresponding to the dashed line, as opposed
to the prediction coming from the Witten--Veneziano relation, depicted as a
solid line. In Fig.~\ref{fig:Fpi} (right) we show the lattice spacing dependence
of these chiral extrapolations, along with the results obtained in
Ref.~\cite{Castellanos:2023rdw}. As we can see, the values of the coupling
considered in both studies are rather large in order for a reliable continuum
extrapolation to be feasible, as cutoff effects might not be
negligible.\footnote{Furthermore, it is worth remarking that the work in
Ref.~\cite{Castellanos:2023rdw} extracted $F_{\pi}$ using an analytic expression
for the residual pion mass in the $\delta$-regime which is in principle valid
only for dimension $D \geq 3$.} Further simulations beyond $\beta=6$ will be
necessary in order to perform a reliable continuum extrapolation, but our
results are in clear tension with those of Ref.~\cite{Castellanos:2023rdw}.

In Fig.~\ref{fig:Gpi} (left) we show the results for the pion matrix
element $G_{\pi}$ for the same three values of the coupling. As suggested by the
scaling obtained in Eq.~(\ref{eq:intro:fpi-gpi-scaling}), we also fit the
simulation data to the functional form
\begin{align}
  \label{eq:schwinger:gpi-fit}
G_{\pi}(m_{R}) = A m_{R}^{1 / 3} + B,
\end{align}
with $A$ and $B$ fitting parameters, finding good agreement. The extrapolated
values at $m_{R}=0$ of these parameters are also shown in Fig.~\ref{fig:Gpi}
as a function of $\beta^{-1 / 2}$. A tentative
linear continuum extrapolation is also displayed, finding a value of $B$ which
agrees with zero.\footnote{Note that for the extrapolation we keep the volume fixed in lattice
  units and only take $\beta \to \infty$, but, since $m_{\pi }L\in[7, 25]$ in our simulations, we do not
expect sizable finite volume effects.} Although more simulations closer to the
continuum would be
suitable, the results seem to validate the picture that the pions {\it dissolve} in the chiral limit~\cite{Georgi:2007ek}, as discussed in
Sec.~\ref{sec:SM}.

\subsection{Nondegenerate case}

\begin{figure*}[t!]
	\centering
	\includegraphics[width=0.49\linewidth,keepaspectratio,angle=0]{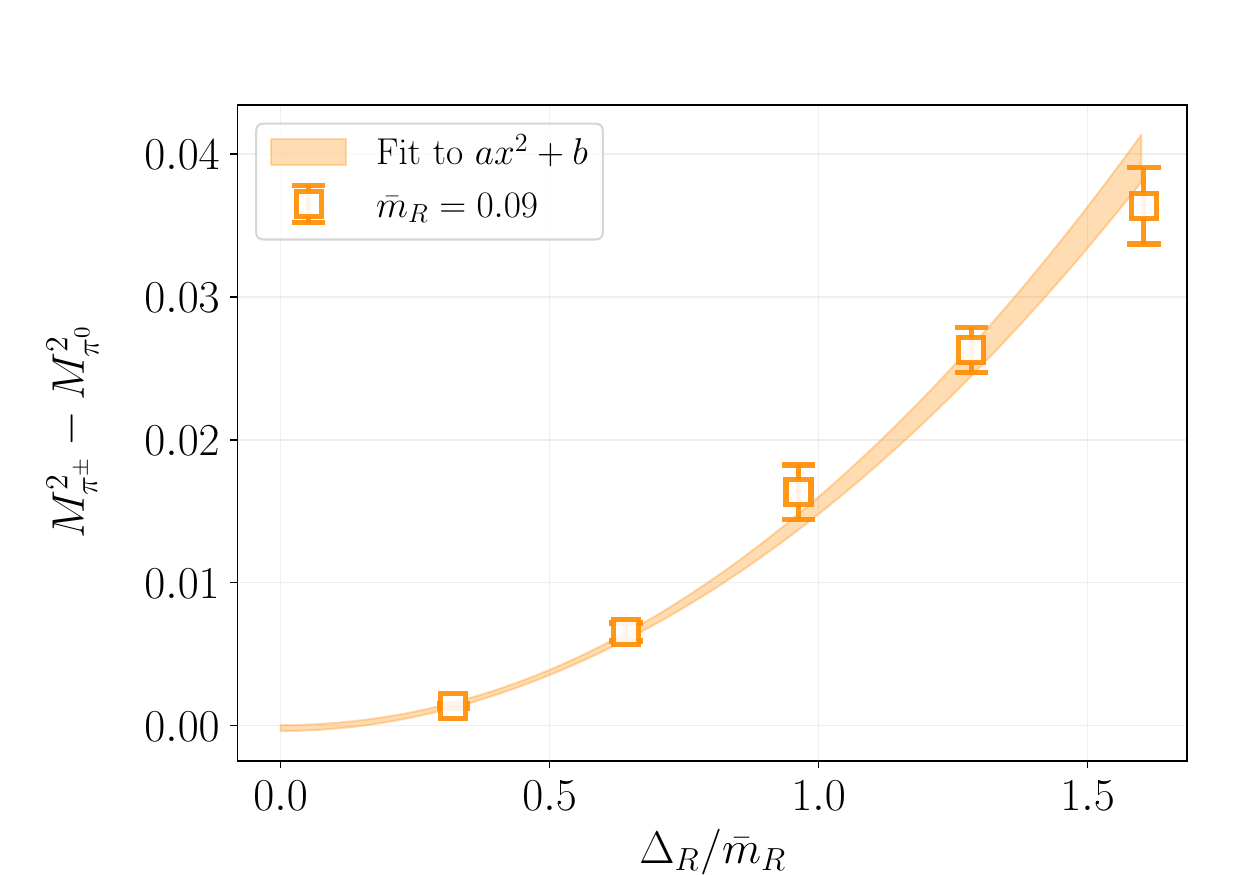}
	\includegraphics[width=0.49\linewidth,keepaspectratio,angle=0]{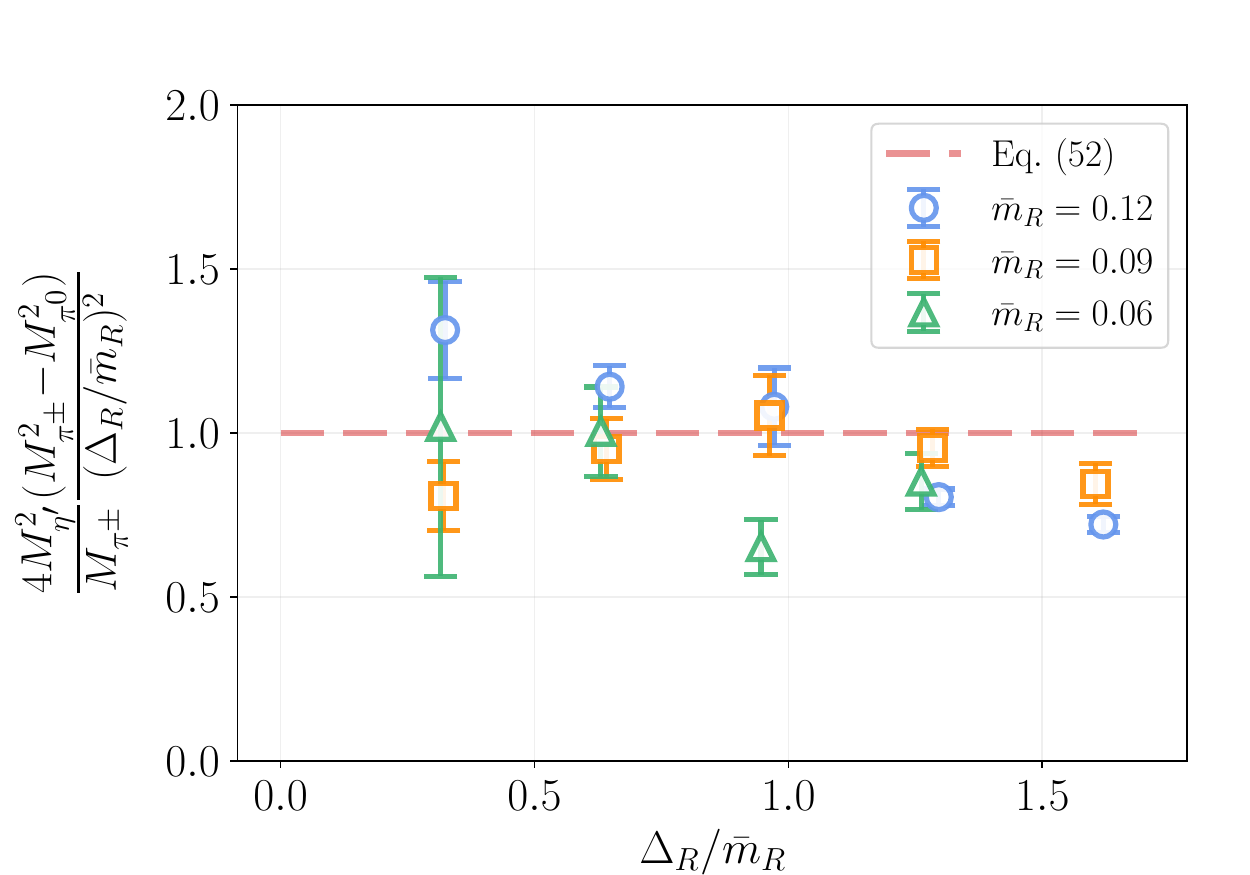}

	\caption{Left: pion mass splitting as a function of the renormalized quark
    mass splitting  $\Delta_{R} = Z_{m} \Delta$
    normalized by the average quark mass $\bar{m}_{R}=0.09$, from simulations at
  $\beta = 4.0$ and $L=64$. The orange band corresponds to a fit to the
  functional form $a x^2 + b$. Right: normalization factor of
  Eq.~(\ref{eq:pion-mass-splitting-result}) as a function of the quark
renormalized mass splitting normalized by the average quark mass, for central
masses $\bar{m}_{R}=\{ 0.06, 0.09, 0.12 \}$ (green triangles, orange squares,
blue circles).}
  \label{schwinger:fig:dm2vsm}
\end{figure*}

\begin{table}[!t]
  \centering
\begin{tabular}{c|c|c|c}
$\bar{m}_{R}$ & $\Delta$ & $M_{\pi^{\pm}}$ & $M_{\pi^{0}}$\\[0pt]
  \hline\hline
0.1203(10) & 0.04 & 0.38650(22) & 0.38019(81)\\[0pt]
 & 0.08 & 0.38535(16) & 0.3629(14)\\[0pt]
 & 0.12 & 0.38335(19) & 0.3335(59)\\[0pt]
 & 0.16 & 0.38045(19) & 0.3121(24)\\[0pt]
 & 0.20 & 0.37610(19) & 0.2741(41)\\[0pt]
 \hline
0.09107(88) & 0.03 & 0.32122(19) & 0.31905(37)\\[0pt]
 & 0.06 & 0.32058(19) & 0.3102(11)\\[0pt]
 & 0.09 & 0.31927(17) & 0.2926(33)\\[0pt]
 & 0.12 & 0.31720(19) & 0.2726(29)\\[0pt]
 & 0.15 & 0.31418(23) & 0.2497(54)\\[0pt]
 \hline
0.06183(75) & 0.02 & 0.24808(20) & 0.24686(60)\\[0pt]
 & 0.04 & 0.24752(18) & 0.24267(70)\\[0pt]
 & 0.06 & 0.24669(20) & 0.23955(96)\\[0pt]
 & 0.08 & 0.24574(22) & 0.2287(18)\\[0pt]
 \hline\hline
\end{tabular}
    \caption{Results of $M_{\pi^{\pm}}$ and $M_{\pi^{0}}$ for the different
      central masses $\bar{m}_{R}$ and bare splittings $\Delta$ at $\beta=4$.}
  \label{tab:IB}
\end{table}

The pion mass splitting in Eq.~(\ref{eq:pion-mass-splitting-result})
can be checked numerically with lattice simulations. To simulate two quark
flavors with different mass we use the RHMC algorithm for a single value of the
coupling, $\beta=4.0$, lattice size $L=64$,
and for three different values for the average renormalized quark
mass: $\bar{m}_{R}=\{ 0.618,\; 0.911,\; 0.120 \}$. For each central value, we
perform simulations for different quark mass splittings in the
range $\Delta_{R} / \bar{m}_{R} \in [0.3,\; 1.7]$,
where $\Delta_{R} \equiv Z_{m} \Delta$ is the renormalized quark mass splitting. To
compute the masses, we use the interpolators
\begin{align}
  \label{eq:latmeth:schwinger-interpolators}
  O_{\pi^{+}}(x) &= \bar{\psi}_{2}(x)\gamma_{5} \psi_{1}(x), \nonumber\\
  O_{\pi^{0}}(x) &= \bar{\psi}_{1}(x)\gamma_{5}\psi_{1}(x) - \bar{\psi}_{2}(x)\gamma_{5}\psi_{2}(x),
\end{align}
and our results are displayed in Table~\ref{tab:IB}.

In Fig.~\ref{schwinger:fig:dm2vsm} (left) we plot the pion mass splitting as a
function of $\Delta_{R} / \bar{m}_{R}$ for $\bar{m}_{R}=0.091$. The results show
good agreement with a fit to a quadratic function, and thus validate the
functional form in Eq.~(\ref{eq:pion-mass-splitting}). In
Fig.~\ref{schwinger:fig:dm2vsm} (right) we show the left-hand side of
Eq.~(\ref{eq:pion-mass-splitting-result}) normalized with the right-hand side,
for the three different central values of the renormalized mass and as a
function of the quark mass splitting. We find good agreement, specially for the
lowest values of the central masses and splittings.

To study the adequacy of the proposed chiral Lagrangian in
Eq.~(\ref{eq:schwinger:chiral-lagrangian}) for small masses more conclusively it
would be good to analyze the masses of the $\eta'$ the $\sigma$ mesons, also
with further simulations for values of the coupling closer to the continuum.
However, the reasonably good agreement of the data with
Eq.~(\ref{eq:pion-mass-splitting}), even for a single value of the coupling,
indicates that the sine-Gordon model resulting from integrating out the $\eta'$
is indeed inadequate to study isospin breaking: from the point of view of the
low-energy pion theory, isospin breaking is a higher-dimensional operator
suppressed with the heavy meson mass.

\section{Conclusions}
\label{sec:conclu}

We have revisited the two-flavor Schwinger model, focusing on two of its most
intriguing features: the existence of a conformal sector in the chiral limit and
the restoration of isospin symmetry in the spectrum in the presence of isospin
breaking.

Although the model is not solvable for nonvanishing fermion masses, predictions
exist in the strong coupling limit, where the light sector of the theory is a
sine-Gordon model. In this limit, there are analytical predictions of various
observables, such as the fermion mass dependence of the pseudoscalar triplet
meson mass and its decay constant. We have confronted these predictions with
lattice simulations of the theory, reaching sufficient statistical precision to
confirm the agreement with the sine-Gordon limit predictions as opposed
to other semiclassical approximations.

We have introduced an effective theory that should describe the light sector of
the theory, based on a chiral effective theory including a dilaton field. In
contrast with the sine-Gordon model, the nonanomalous flavor symmetry is
explicit, while scale invariance is recovered in the massless limit. The
effective theory reproduces the correct fermion mass dependence of the
pseudoscalar meson mass, the scalar-to-pseudoscalar meson mass ratio---which
is $\sqrt{3}$---as well as the isospin symmetric spectrum in the presence of
nondegenerate masses. Furthermore, if the pseudoscalar singlet meson is added
to the effective theory so as to reproduce the $U(1)_A$ anomaly Ward identity,
a parameter-free prediction for the splitting of the isospin triplet is derived.

Finally, we have studied the triplet pseudoscalar masses with the presence of
isospin breaking from nondegenerate fermion masses, finding agreement with the
expectation based on the effective theory---see Fig.~\ref{schwinger:fig:dm2vsm}.
The concept of automatic fine-tuning of isospin, introduced in
Ref.~\cite{Georgi:2020jik}, is discussed and reinterpreted as a decoupling
effect of the $\eta'$.

\section*{Acknowledgments}

We thank J.~Baeza-Ballesteros, A.~Ramos, L.~Del Debbio and A.~Donini for useful
discussions.  Our activities are partially funded by the Staff Exchange Grant
Agreement No. 101086085- ASYMMETRY, by the Spanish Ministerio de Ciencia e
Innovaci\'on project PID2020-113644GB-I00 and PID2023-148162NB-C21, and by
Generalitat Valenciana through the Grants No. CIPROM/2022/69 and No.
PROMETEO/2021/083-03.

\bibliography{biblio.bib}

\end{document}